\documentclass[structabstract]{aa}  
\usepackage{graphicx,amsmath}
\usepackage{txfonts}
\usepackage{natbib}
\newcommand{\xr}[1]{\xrightarrow{#1}}
\newcommand{\tx}[1]{\mathrm{#1}}
\begin{document}
   \title{The role of OH in the chemical evolution of protoplanetary disks\\
          II. Gas-rich environments}

   \author{G. Chaparro Molano
          \and
          I. Kamp
          }

   \institute{Kapteyn Astronomical Institute, Postbus 800, 9747 AV Groningen, The Netherlands
             }

   \date{2012}

 \abstract{We present a method for including gas extinction of cosmic-ray-generated UV photons in chemical models of the midplane of protoplanetary disks, focusing on its implications on ice formation and chemical evolution.}
{Our goal is to improve on chemical models by treating cosmic rays, the main source of ionization in the midplane of the disk, in a way that is consistent with current knowledge of the gas and grain environment present in those regions. We trace the effects of cosmic rays by identifying the main chemical reaction channels and also the main contributors to the gas opacity to cosmic-ray-induced UV photons. This information is crucial in implementing gas opacities for cosmic-ray-induced reactions in full 2D protoplanetary disk models. }
{We considered time-dependent chemical models within the range 1-10 AU in the midplane of a T Tauri disk. The extinction of cosmic-ray-induced UV photons by gaseous species was included in the calculation of photorates at each timestep. We integrated the ionization and dissociation cross sections of all atoms/molecules over the cosmic-ray-induced UV emission spectrum of H$_2$. By analyzing the relative contribution of each gas phase species over time, we were able to identify the main contributors to the gas opacity in the midplane of protoplanetary disks.}
{At 1 AU the gas opacity contributes up to 28.2\% of the total opacity, including the dust contribution. At 3-5 AU the gas contribution is 14.5\% of the total opacity, and at 7-8 AU it reaches a value of 12.2\%. As expected, at 10-15 AU freeze-out of species causes the gas contribution to the total opacity to be very low (6\%). The main contributors to the gas opacity are CO, CO$_2$, S, SiO, and O$_2$. OH also contributes to the gas opacity, but only at 10-15 AU.}{} 

   \keywords{Astrochemistry; Protoplanetary disks; Molecular processes; (ISM:) cosmic rays}

   \maketitle
%

\section{Introduction}

The midplane of protoplanetary disks has been considered a dead zone, because the lack of a source of ionization prevents the development of magneto-rotational instabilities, which are thought to drive the accretion process. The midplane of a disk corresponding to a Class II source around a T Tauri star is opaque to stellar and interstellar UV \citep{zadel,wkt} and X-ray \citep{glassgold,giamba1} photons, which corresponds to a region located at $z/r$$<$0.05 and $1$$<$$r$$<$$10$ AU. However, most regions of the midplane are far from being dead zones because of cosmic rays, which directly ionize the gas and heat the dust grains. An important consequence of the interaction of cosmic rays and H$_2$ molecules is the emission of a locally generated UV field that can ionize/dissociate species. More specifically, regions that have a value for $A_V$ of a few are dominated by cosmic-ray processing, since they can penetrate column densities of $\Sigma$ $\sim$ 150 g cm$^{-2}$ \citep{umebayashi}. \\ 

Hence, the midplane of the disk between 1 and 10 AU can be viewed as a cosmic-ray dominated region, where cosmic rays are the main source of ionization and therefore the main driver of the chemical evolution. Steady-state chemical models applied to these regions cannot fully describe the chemical evolution of the midplane, as the chemical relaxation timescale can be as long as 10$^8$ yr \citep{wkt}, which is longer than the lifetime of the disk \citep{fedele,haisch,wiebe}. For this reason it is necessary to approach the study of the chemical evolution of the disk from a time-dependent model. One of the main catalysts of the cosmic-ray driven chemistry in these regions is OH \citep{paper1}, especially for the formation of CO and H$_2$O. \\

Current chemical models of protoplanetary disks include to a large extent the effects of cosmic rays, whether they model the chemistry using steady-state \citep{wkt,will,gorti,thi} or time-dependent \citep{visser,visser2,walsh,semenov} solutions. However, the effects of the local gas opacity and grain growth in cosmic-ray-induced UV processes are largely overlooked, as the parameters for estimating cosmic-ray-induced photoionization/dissociation rates are usually taken from molecular cloud literature \citep{gldh,umist}.\\ 

In molecular cloud modeling \citep{cecchi,shenvd} it is customary to ignore those effects, as they are not relevant for the cold, gas-poor environment present deep inside molecular clouds and the interstellar medium (ISM). Another important factor to consider is grain growth in protoplanetary disks, which reduces dust UV opacity compared to molecular clouds\footnote{See \cite{dalessio} for evidence of grain growth from spectral energy distributions of protoplanetary disks}. This leads to an enhancement in the cosmic-ray-induced UV flux \citep{paper1} with respect to previous ISM based estimates \citep{shenvd}, especially in gas-poor regions. This field, which is enhanced by a factor 40 at 10 AU, drives gas phase formation pathways for saturated molecules that can later freeze on the dust surface. Wherever the physical conditions allow for penetration of cosmic rays in disks, our analysis of cosmic-ray driven chemistry applies. Since we aim to perform a quantitative analysis, we chose the physical conditions from a particular disk model. However, our results do not depend on that particular choice.\\

This paper is structured as follows. The strategy to implement our model in Sect. 2 is followed by a discussion of the physical conditions in gas-rich regions of the disk midplane in Sect. 3. In Sect. 4 we discuss our treatment of cosmic-ray-induced UV photoprocesses including the effects of gas opacity. Sect. 5 deals with the particulars of our chemical model, followed by a summary (Sect. 6) and a discussion (Sect. 7) of our results. Finally, the main conclusions from our work are summarized in Sect. 8.
\begin{table}
\begin{center}
\caption{Distance from the star, temperature, and density conditions corresponding to midplane regions in the protoplanetary disk structure in Fig. \ref{prod}, following the \textsc{ProDiMo} simulation of a passive irradiated disk \citep{wkt}. The stellar parameters used in this simulation are found in Table \ref{tabpar}.}
\label{tabpoints}\renewcommand{\arraystretch}{1.5}
\begin{tabular}{ccc}\hline\hline
 $r$ (AU) & $T$ (K) & $n_\mathrm{\langle H\rangle}$ ($\mathrm{cm^{-3}}$)
 \\ \hline
 1 & 80 & $10^{14}$ \\ 
 3-4 & 65 & $10^{12}$ \\ 
 7-8 & 40 & $10^{11}$ \\ 
 10-15 & 20 & $10^{10}$ \\ \hline
\end{tabular}
\end{center}
\end{table}
\section{Methodology}

Our goal is to identify the main chemical contributors to the extinction of cosmic-ray-induced UV (CRUV) photons in different regions of the midplane of a T Tauri disk. The temperature in the disk midplane can be high enough to prevent species from freezing onto the surface of grains immediately after their gas phase formation. This general freeze-out of chemical species is found to happen at distances beyond approximately 10 AU \citep{paper1}. Thus, we chose the 1-10 AU range of the disk midplane in this work, becuase the temperature and density conditions are ideal for studying regions with very different gas compositions. For example, the temperature at 7-8 AU (see Table \ref{tabpoints}) coincides with the onset of thermal desorption of CO, which will evaporate from the surface of grains while leaving the abundances of other frozen species for the most part unchanged.\\

The physical input conditions for our chemical evolution models are given by a \textsc{ProDiMo} model \citep{wkt}, and are listed in Tables \ref{tabpoints} and \ref{tabpar}. We implemented a time-dependent calculation of the CRUV photorates that includes the extinction provided by the dust and also by gas species, which depends on their abundance. For this reason, we recomputed the CRUV photorate at each timestep of the simulation to account for the changes in gas phase abundances. This extinction was then integrated over the wavelength range and emission probability of CRUV photons. The emission probability was obtained from Lyman and Werner emission lines of H$_2$ that is excited either by direct cosmic-ray interactions or by secondary electrons generated in cosmic-ray excitation of H$_2$ \citep{prasad}. \\

Using this scheme, we can trace species that have a strong impact on the absorption of CRUV photons for the midplane of T Tauri disks. By taking into account the extinction contribution of dust grains, we can compare it to the gas extinction, thus obtaining a time-dependent value for the opacities over the CRUV wavelength range. We implemented this method in our time-dependent chemical rate equation solver \verb"chem_compact", which we previously used for studying ice formation in the comet formation zone \citep{paper1} of a passive T Tauri disk with a low accretion rate. This code was benchmarked against steady-state chemical abundances from the \textsc{ProDiMo} simulation from \cite{wkt}.

\begin{figure}
\begin{center}
\includegraphics[scale=0.52]{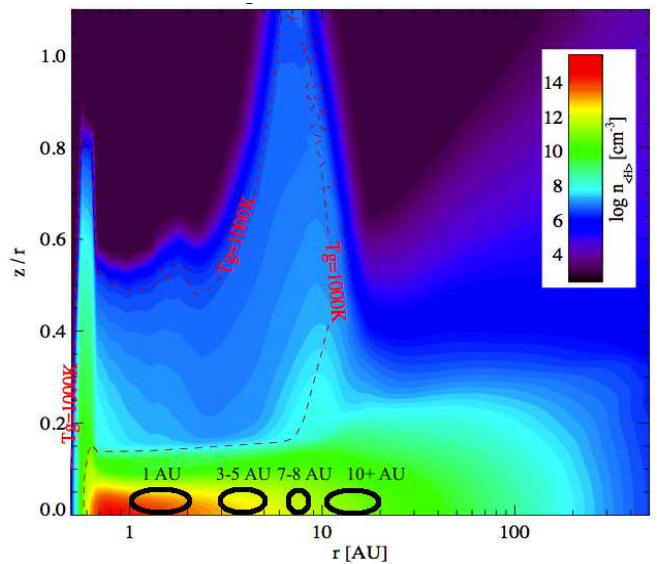}
\caption{Density structure model of a T Tauri disk as a function of radial distance from the star and the relative height, following the \textsc{ProDiMo} simulation \citep{wkt}. The black ovals show the regions of the disk according to Table \ref{tabpoints}. The relevant disk parameters are provided in Table \ref{tabpar}.}\label{prod}
\end{center}
\end{figure}

\subsection{Gas-rich regions}

\begin{table}
\begin{center}
\caption{Table of modeling parameters.}
\label{tabpar}\renewcommand{\arraystretch}{1.1}
\begin{tabular}{lcc}\hline\hline
Parameter & Symbol & Value \\ \hline
Stellar mass & $M_*$ & 1 $M_\odot$ \\
Effective temperature & $T_\mathrm{eff}$ & 5770 K \\
Stellar luminosity & $L_*$ & 1 $L_\odot$ \\ \hline
Disk mass & $M_\mathrm{D}$ & 0.01$M_\odot$\\ 
Inner disk radius & $R_\mathrm{in}$ & 0.5 AU\\
Outer disk radius & $R_\mathrm{out}$ & 500 AU\\
Gas surface density power law index & $\epsilon$ & 1.5\\
Dust-to-gas mass ratio & $\rho_d/\rho_g$ & 0.01\\ 
Minimum dust grain size & $a_\mathrm{min}$ & 0.1\,$\mu$m\\ 
Maximum dust grain size & $a_\mathrm{max}$ & 10\,$\mu$m\\ 
Mean molecular weight & $\mu$ & 1.35\\
Dust grain size power law index & $p$ & 3.5\\ 
Dust material mass density & $\rho_\mathrm{gmd}$ & 2.5\,g$\cdot$cm$^{-3}$\\ 
Dust grain albedo (UV) & $\omega$ & 0.57\\ 
Dust opacity (UV) & $\kappa_\mathrm{UV}$ & 6.8$\times10^3$\,cm$^2\cdot$g$^{-1}$\\ 
Cosmic ray ionization rate (H$_2$) & $\zeta_\mathrm{H_2}$ & $5\times10^{-17}\,\mathrm{s^{-1}}$\\ \hline
Number of active layers & $N_\mathrm{Lay}$ & 2\\ 
Adsorption site area & $A_\mathrm{site}$ & $6.67\times10^{-16}$\,cm$^{2}$\\ \hline
C adsorption energy & $E_b^\mathrm{C}$ & 630 K \\
CO adsorption energy & $E_b^\mathrm{CO}$ & 960 K \\
CO$_2$ adsorption energy & $E_b^\mathrm{CO_2}$ & 2000 K \\
CH$_3$ adsorption energy & $E_b^\mathrm{CH_3}$ & 920 K \\
CH$_4$ adsorption energy & $E_b^\mathrm{CH_4}$ & 1100 K \\
O adsorption energy & $E_b^\mathrm{O}$ & 630 K \\
O$_2$ adsorption energy & $E_b^\mathrm{O_2}$ & 960 K \\
OH adsorption energy & $E_b^\mathrm{OH}$ & 1000 K \\
H$_2$O adsorption energy & $E_b^\mathrm{O}$ & 4800 K \\
Si adsorption energy & $E_b^\mathrm{Si}$ & 2100 K \\
SiH adsorption energy & $E_b^\mathrm{SiH}$ & 2300 K \\
SiO adsorption energy & $E_b^\mathrm{SiO}$ & 2800 K \\
Fe adsorption energy & $E_b^\mathrm{Fe}$ & 3300 K \\
Mg adsorption energy & $E_b^\mathrm{Mg}$ & 4200 K \\ \hline
\end{tabular}
\end{center}
\end{table}

Cosmic ray-induced UV photons can be absorbed by the material in the local environment where they are generated. Both gas and dust can absorb these photons and become a source of local extinction, but this depends on the local density and temperature conditions. For instance, at 10 AU most of the material is frozen onto the surface of dust grains, which means that the extinction of CRUV depends entirely on the local dust properties. In the absence of gas extinction, grain aggregation in protoplanetary disks can lead to an enhanced CRUV flux \citep{paper1}. In these gas-poor environments CRUV photoprocesses will not be affected by the composition of the gas and the chemistry can be described in a fairly straightforward fashion.\\

By contrast, in regions closer to the central star the environment is heated up and most of the chemical species will stay in the gas phase while leaving significantly reduced layers of frozen species. Any change in the chemical composition of the gas will either enhance the CRUV field or quench it efficiently, depending on the CRUV cross section of the dominant species in the gas. Thus, if we aim to understand the complex coupling effects between chemistry and CRUV photons, we need to fully incorporate the contribution of the gas in the local CRUV extinction. \\

In Fig. \ref{prod} we show the regions under study in a plot of the density structure of the disk obtained using \textsc{ProDiMo} \citep{wkt}, which provides us with a self-consistent hydrostatic structure from which we obtain parameters such as temperature, density, and intensity of the local UV field compared to the ISM Draine field. By studying regions at different distances from the star, we can identify the species that are locally dominant in their CRUV opacity.\\

The specific temperature and density conditions for each specific region of the disk midplane are found in Table \ref{tabpoints}. Near the inner rim (at 1 AU from the star) we focus on the role of CRUV photoprocesses in the highly efficient OH forming region near the inner rim. Moving farther away from the star, the temperature and the density decrease, which causes more material to freeze onto the surface of dust grains. To understand the role of freeze-out, we probed the disk at two more regions: at 3-4 AU and 7-8 AU. For the 10 AU region we refer the reader to part I of our paper \citep{paper1}. 

\section{Cosmic-ray-induced processes}\label{crip}

Cosmic rays penetrate to the disk midplane predominantly from the vertical direction because the column density along all other directions is too high. From the analysis of cosmic-ray penetration in \cite{semcr}, it follows that in the midplane of the particular generic T Tauri disk chosen in this study (see Table \ref{tabpar} for a list of parameters), at radial distances larger than 1 AU cosmic rays can penetrate almost unhindered. Recent work by \cite{padov} on the penetration of cosmic rays in molecular clouds suggests that interaction with magnetic fields is more important than previously thought. However, those results do not necessarily apply for the particular magnetic field geometry of protoplanetary disks, and more detailed modeling is necessary to clear the picture of cosmic-ray and magnetic field interactions.  \\

The main chemical byproducts of direct cosmic-ray ionization are H$_3^+$ (from H$_2^+$) and He$^+$ \citep{klemperer}. While He$^+$ is very good at dissociating molecules and passing on its charge to the products of the reaction, H$_3^+$ hydrogenates and ionizes CH compounds, and helps create water from atomic oxygen \citep{paper1}.\\

Cosmic rays can also ionize the medium in a more subtle way: by inducing a UV field that comes from secondary ionization of molecular hydrogen. The process, known as the Prasad Tarafdar mechanism \citep{prasad}, starts when an electron with a typical energy of 30 eV is released after the cosmic-ray ionization of an H$_2$ molecule. This secondary electron can also ionize another H$_2$ molecule. A value for the total rate $\zeta_\mathrm{H_2}$ of both direct and secondary ionization of H$_2$ is not entirely agreed upon, but a conservative value of
\begin{equation}\label{zeta}
  \zeta_\mathrm{H_2}=5\times10^{-17}\ \,\mathrm{s}^{-1}\ 
\end{equation} 
has been obtained both for H$_3^+$ measurements in the ISM \citep{indriolo} and from theoretical estimations based on measured cosmic-ray spectra \citep{cecchi,micel}.\\

The emitted electron then hits another (neutral) H$_2$ molecule, which leaves it in an excited electronic state \citep{sdl,riahi}, after which it spontaneously decays to the excited vibrational states of the $B$ $^1\Sigma_u^+$ and $C$ $^1\Sigma_u$ levels. In the subsequent decay to excited vibrational states of the ground electronic level $X$ $^1\Sigma^+_g$ of H$_2$, Lyman and Werner photons are emitted in the 90-170 nm range. These cosmic-ray-induced UV photons can then either ionize/dissociate a gas species or hit a dust grain.\\

We define the CRUV photoprocess efficiency as the fraction of CRUV photons that dissociate a species and are not locally absorbed by the gas or dust:
\begin{equation}\label{fullint}
 \gamma_i=\int_{1.76\,\mathrm{PHz}}^{3.28\,\mathrm{PHz}}\frac{P(\nu)\sigma_i(\nu)}{\sigma_{\mathrm{tot}}(\nu)}\,d\nu\ .
\end{equation} 
Here $P$ is the emission probability profile of a CRUV photon, $\sigma_i$ is the photoprocess cross section (in cm$^2$ per species), and $\sigma_{\mathrm{tot}}$ is the total (gas+dust) cross section, which is a measure of the local extinction. If a given species with a high CRUV cross section is very abundant in the gas phase, the photo rate of that species will have a maximum value and will be low for all other species in the gas. This shielding effect cannot be ignored in regions where high-density/temperature combinations create a gas-rich environment.\\

Finally, cosmic rays can directly heat dust grains and cause desorption of ices. We took this effect into account, but it is more predominant in outer regions of the disk.

\subsection{CRUV emission probability profile}\label{crepp}

The emission probability profile of CRUV is obtained from the transition probability of the first three electronic levels ($B$ $^1\Sigma_u^+$ and $C$ $^1\Sigma_u$) of molecular hydrogen. The cross section for excitation of H$_2$ into a level $v'$ is proportional to the optical band oscillator strength $f_{0v'}$.
\begin{equation}
 \sigma_{v'0}\propto f_{0v'}\ .
\end{equation} 
We use proportionality here, as it is enough to obtain a normalized emission probability profile. The oscillator strength is
\begin{equation}
 f_{0v'}\propto A_{v'0}\frac{g_{v'}}{g_0}\frac{1}{\nu_{v'0}^2}\ .
\end{equation} 
Here $g$ is the statistical weight. The emission probability for a transition from the level $v'J'$ of an electronically excited state $i$ to the $v'J'$ level of the ground electronic state is then proportional to the Einstein $A$ coefficient for emission and the cross section for excitation:
\begin{equation}
 p(\nu_{v'J'}^i)\propto\sigma^i_{v'0}A^i_{v'J',v''J''}=f^i_{0v'}A^i_{v'J',v''J''}\ .
\end{equation} 
This probability is normalized over all transitions to $v''J''$ levels of the ground electronic state:
\begin{equation}
 P(\nu_{v'J'}^i)=\frac{f^i_{0v'}A^i_{v'J',v''J''}}{\sum_{v'J'}\sum_{v''J''}f^i_{0v'}A^i_{v'J',v''J''}}\ .
\end{equation} 
Each transition probability is then convolved into a Voigt line profile to account for both natural and thermal broadening:
\begin{equation}
 P(\nu;\nu_{v'J'}^i)=P(\nu_{v'J'}^i)\phi_\mathrm{V}(\nu-\nu_{v'J'}^i)\ .
\end{equation} 
The emission probability profile for each excited electronic state is then
\begin{equation}
 P^i(\nu)=\sum_{v'J'}P(\nu;\nu_{v'J'}^i)\ .
\end{equation} 
Thus, the probability that a CRUV photon will be emitted after an H$_2$ cosmic-ray ionization process is
\begin{equation}
 P(\nu)=P^{B\,^1\Sigma_u^+}(\nu)+P^{C\,^1\Sigma_u^+}(\nu)+P^{C\,^1\Sigma_u^-}(\nu)
\end{equation} 
This probability is normalized over the frequency range $1.76-3.28\,\mathrm{PHz}$ (90-170 nm). We obtained the Einstein $A$ coefficients and the frequencies for the relevant transitions from the tables of \cite{abgrall}, and the statistical weights were calculated from the guidelines in Appendix A of \cite{plasma}.

\subsection{CRUV gas opacity}

The total cross section $\sigma_{\mathrm{tot}}$ in Eq. (\ref{fullint}) is (in cm$^2$ per hydrogen atom)
\begin{equation}\label{crosssect}
 \sigma_{\mathrm{tot}}(\nu)=\tilde{\sigma}_\mathrm{\langle H\rangle}^\mathrm{dust}(1-\omega)+\sum_{j}\xi_j\sigma_j(\nu)\ .
\end{equation}
Here $\omega$ is the grain albedo, $\xi_j={n_j}/{n_\mathrm{\langle H\rangle}}$ is the abundance of the species $j$, and $\tilde{\sigma}_\mathrm{\langle H\rangle}^\mathrm{dust}$ is the grain UV extinction cross section per hydrogen atom. We refer the reader to Eq. (10) in \cite{paper1} to see how the cross section relates to the dust UV opacity and other dust parameters in Table \ref{tabpar}. Since the dust UV extinction curve is fairly flat in the CRUV frequency range, we used a frequency average for this value. Due to photon conservation, the previous expression leads to the following relation:
\begin{equation}
 \tilde{\sigma}_\mathrm{\langle H\rangle}^\mathrm{dust}(1-\omega)\gamma_\tx{dust}+\xi_i\gamma_i=1\ .
\end{equation} 
Here $\gamma_\tx{dust}$ is
\begin{equation}
\gamma_\tx{dust}=\int\frac{P(\nu)}{\sigma_{\mathrm{tot}}(\nu)}\,d\nu\ .
\end{equation} 
Thus $\xi_i\gamma_i$ measures the fractional contribution from the species $i$ to the CRUV extinction. The shape of the radiation field that dissociates or ionizes a species $F(\nu)$ not only depends on the CRUV emission probability, but also on the CRUV extinction of all other species $\sigma_{\mathrm{tot}}(\nu)$:
\begin{equation}\label{eqf}
 F(\nu)=\frac{P(\nu)}{\sigma_{\mathrm{tot}}(\nu)}\ .
\end{equation} 
The frequency-dependent opacity for a species $i$ can be written as
\begin{equation}\label{eqkap}
 \kappa_i(\nu)=\xi_i\,\sigma_i(\nu)\,\frac{n_\mathrm{\langle H\rangle}}{\rho_\tx{gas}}\ \mathrm{cm^2\ g^{-1}\ (gas)}\ .
\end{equation} 
Here $\rho_\tx{gas}$=$n_\mathrm{\langle H\rangle}\,\mu\,m_\tx{H}$, with $\mu$ being the mean molecular weight (see Table \ref{tabpar}). We can now define a ``gray'' (frequency averaged) opacity as
\begin{equation}
 \langle\kappa_i\rangle=\frac{\int_{1.76\,\mathrm{PHz}}^{3.28\,\mathrm{PHz}} F(\nu)\kappa_i(\nu)\,d\nu}{\int_{1.76\,\mathrm{PHz}}^{3.28\,\mathrm{PHz}} F(\nu)\,d\nu}\ .
\end{equation} 
Using Eqs. (\ref{eqf}) and (\ref{eqkap}) this expression can be rewritten as
\begin{equation}\label{opacity}
 \langle\kappa_i\rangle=\frac{\xi_i\,\gamma_i}{\gamma_\tx{dust}\,\mu\,m_\tx{H}}\ .
\end{equation} 
With this expression, we can accurately measure the opacity of each species throughout a time-dependent chemistry run. The variables $\xi_i$, $\gamma_i$ and $\gamma_\tx{dust}$ are calculated at every timestep of the simulation as part of the CRUV photoprocess rate calculation. Given that the CRUV photoprocess rate constant $k_i\propto\gamma_i$, the resulting rate is then coupled to the abundance of all other species that contribute to the opacity.\\
\begin{table}
\begin{center}
\caption{Table of chemical species in the chemical networks. \# indicates an ice species.}\label{tabspe}
\renewcommand{\arraystretch}{1.1}
\begin{tabular}{cc}\hline\hline
Type & Symbol  \\ \hline
Atoms & H, He, C, O, S, Si, Mg, Fe \\ \hline
Ions & He$^+$, Si$^+$, Fe$^+$, H$^-$, H$^+$, C$^+$, \\
 & O$^+$, S$^+$, Mg$^+$ \\ \hline
Molecules & H$_2$, H$_2$O, CH$_2$, HCO, SiO, CO$_2$, \\ 
 & SiH, CH$_3$, CH$_4$, OH, O$_2$, CO, CH, H$_2$CO  \\  \hline
Molecular & HCO$^+$, CH$_2^+$, H$_3^+$, SiH$^+$, SiO$^+$, \\
Ions &  CH$_4^+$, H$_3$O$^+$, H$_3$O$^+$, SiH$_2^+$, CH$_5^+$, \\ 
 &  CH$_3^+$, H$_2$O$^+$, SiOH$^+$, CH$^+$, H$_2^+$, \\ 
 & O$_2^+$, CO$^+$, OH$^+$, CO$_2^+$ \\ \hline
Ice &  C\#, CO\#, CO$_2$\#, CH$_3$\#, CH$_4$\#, \\
 & O\#, O$_2$\#, OH\#, H$_2$O\#, \\
 & Si\#, SiH\#, SiO\#, Fe\#, Mg\# \\ \hline
\end{tabular}
\end{center}
\end{table}

Since photoionization/dissociation of species can be continuum and/or line processes, we convolved the line photoprocess cross sections with a Voigt profile in order for both natural and Doppler line broadening. The total cross section is then the sum of the line and continuum processes for each molecule. Values for the cross sections were obtained from the tables in the Leiden photoprocess database hosted by E. van Dishoeck \verb"http://www.strw.leidenuniv.nl/~ewine/photo/" \citep{faraday}.\\

It is safe to consider only the effects of dust CRUV extinction if we know \textit{a priori} that most of the gas species are frozen onto the surface of grains. This means that if gas phase abundances are low enough to be neglected in Eq. (\ref{crosssect}), dust grains will be the sole contributor to opacity. This approach is frequently taken in models that include CRUV photoprocesses \citep{sdl,gld,gldh,umist}, mostly because it is valid for ISM and molecular cloud conditions (low molecular gas abundances).\\

At 10 AU the environment is cold enough ($T_\mathrm{gas}$=20 K, $n_\mathrm{\langle H \rangle}$=10$^{10}$ cm$^{-3}$) to study the chemical evolution without taking into account the gas opacity \citep{paper1}. Under these conditions, the CRUV rate constant takes a simplified form that does not depend on species abundances.

\section{Chemical model} \label{cm}

The code \verb"chem_compact", described in our previous paper \citep{paper1}, is our VODE based \citep{vode} gas/grain chemical rate equation solver. In it we include a reaction network based on the \textsc{Umist}06 database for astrochemistry \citep{umist} including H$_2$ formation on grains \citep{cazaux} and ad/desorption reactions: Adsorption and thermal and stellar UV photodesorption from \cite{aikawa,leger,oberg}, cosmic-ray direct desorption from \cite{herbst} and cosmic-ray-induced photodesorption from \cite{roberts}. Surface reactions are not considered because they are beyond the scope of this paper.\\

Table \ref{tabspe} lists all gas and ice species considered in our model. In this work we use the low metal initial abundances from \cite{jenkins}, where absorption lines of these metals are measured\footnote{Despite assuming significant metal depletion, the metal abundances here are about a factor 10 higher than in \cite{graedel} or \cite{lee2}.} from various clouds against a bright background star. The low abundances imply that Si, Fe, and Mg condense into dust grains before the formation of the disk. This is particularly relevant for the formation of SiO and its maximum abundance levels, although formation of CO and H$_2$O is not affected.\\

Initial conditions for our disk model were obtained by running our chemical evolution code under molecular cloud conditions ($T=20$ K, $n_\mathrm{\langle H\rangle}=10^{6}$ cm$^{-3}$) from atomic low-metallicity abundances. The resulting abundances after 10$^7$ yr were used as initial conditions. Table \ref{tababu} lists the initial abundances for the molecular cloud run (atomic) and those for the disk model. Atomic abundances (column 2 in Table \ref{tababu}) were used as input for the molecular cloud run, which yielded the molecular abundances (column 4 in Table \ref{tababu}) that we took as initial abundances for our disk model. \\

We decided to ignore sulphur chemistry beyond ionization of atomic S because its chemical network is only weakly coupled to CO formation and is decoupled from other species considered here \citep{sd}. Atomic sulphur was included as it contributes to the metallicity, and its ionization can be an important source of charge exchange for other species. By ignoring other S-bearing species, we can still arrive at an upper limit for the importance of atomic S in the CRUV gas opacity. \\

To estimate the effects of gas opacity in the CRUV photoprocess rates and therefore in the chemical evolution, we replaced the relevant rates from the \textsc{umist} \citep{umist} database with our own rates (Sect. \ref{crip}). As these rates depend on $\xi_i$, we recalculated them at every timestep, using the new values for the gas abundances. This means that at every timestep, the integral in Eq. (\ref{fullint}) needs to be evaluated from the previous timestep. 

\section{Results}
\begin{table}
\begin{center}
\caption{Table of initial abundances, atomic (column 2) and molecular (column 3). \# indicates an ice species.}\label{tababu}
\renewcommand{\arraystretch}{1.1}
\begin{tabular}{cc|cc}\hline\hline
Species & Abundance & Species & Abundance \\ 
(Atomic) & $\xi_i$ & (Molecular) & $\xi_i$ \\ \hline
H	& 1                    & H$_2$ &	5.00$\times10^{-1}$ \\
He	& 8.51$\times10^{-2}$  & He &	8.51$\times10^{-2}$ \\ 
O	& 3.31$\times10^{-4}$  & CO &	1.76$\times10^{-4}$ \\  
C	& 1.78$\times10^{-4}$  & H$_2$O\# &	1.09$\times10^{-4}$ \\ 
Fe	& 5.01$\times10^{-6}$  & O &	3.55$\times10^{-5}$ \\   
Mg	& 2.24$\times10^{-6}$  & OH &	5.66$\times10^{-6}$  \\  
S	& 1.90$\times10^{-6}$  & Fe\# &	5.01$\times10^{-6}$   \\ 
Si	& 1.78$\times10^{-6}$  & H &	2.56$\times10^{-6}$    \\  
         & &                    H$_2$O &	2.55$\times10^{-6}$\\
         & &                    Mg\# &	2.24$\times10^{-6}$\\
         & &                    S &	1.73$\times10^{-6}$\\
         & &                    H$_2$CO\# &	1.63$\times10^{-6}$\\
         & &                    SiO\# &	1.02$\times10^{-6}$\\ \hline
\end{tabular}             
\end{center}              
\end{table}    

\begin{table}
\begin{center}
\caption{Comparison between the gas and dust opacity:average photoprocess cross section (column 2), maximum gas phase abundance required to produce an opacity equal to the dust opacity for ISM type dust (column 3) and for protoplanetary disk type dust (column 4).}
\label{tabgas}\renewcommand{\arraystretch}{1.2}
\begin{tabular}{cccc}\hline\hline		
Species	& $\bar{\sigma}_i$ (cm$^2$)	& Req. $n_i/n_\mathrm{\langle H\rangle}$ & Req. $n_i/n_\mathrm{\langle H\rangle}$  \\
(i) & (90-170 nm) & (PPD) & (\textsc{ISM}) \\ \hline
C	& 3.85$\times10^{-18}$	& 3.98$\times10^{-5}$ & 5.20$\times10^{-4}$ \\
CH	& 6.61$\times10^{-18}$	& 2.31$\times10^{-5}$ & 1.68$\times10^{-5}$ \\
CH$^+$	& 4.86$\times10^{-18}$	& 3.15$\times10^{-5}$ & 4.12$\times10^{-4}$ \\
CH$_2$	& 1.50$\times10^{-17}$	& 1.02$\times10^{-5}$ & 9.47$\times10^{-5}$ \\
CH$_3$	& 1.32$\times10^{-16}$	& 1.16$\times10^{-6}$ & 1.52$\times10^{-5}$ \\
CH$_4$	& 6.13$\times10^{-16}$	& 2.50$\times10^{-7}$ & 3.26$\times10^{-6}$ \\
OH	& 7.82$\times10^{-18}$	& 1.96$\times10^{-5}$ & 2.56$\times10^{-4}$ \\
H$_2$O	& 4.00$\times10^{-18}$	& 3.83$\times10^{-5}$ & 2.87$\times10^{-4}$ \\
Fe	& 1.80$\times10^{-18}$	& 8.49$\times10^{-5}$ & 1.11$\times10^{-3}$ \\
Mg	& 9.20$\times10^{-19}$	& 1.66$\times10^{-4}$ & 4.12$\times10^{-4}$ \\
CO	& 1.82$\times10^{-17}$	& 8.40$\times10^{-6}$ & 1.10$\times10^{-4}$ \\
Si	& 2.14$\times10^{-17}$	& 7.14$\times10^{-6}$ & 9.33$\times10^{-5}$ \\
HCO	& 2.99$\times10^{-16}$	& 5.12$\times10^{-7}$ & 6.69$\times10^{-6}$ \\
SiH	& 6.35$\times10^{-17}$	& 2.41$\times10^{-6}$ & 3.15$\times10^{-5}$ \\
H$_2$CO	& 3.58$\times10^{-16}$	& 4.27$\times10^{-7}$ & 5.59$\times10^{-6}$ \\
O$_2$	& 7.20$\times10^{-18}$	& 2.13$\times10^{-5}$ & 5.77$\times10^{-5}$ \\
S	& 3.36$\times10^{-18}$	& 4.55$\times10^{-5}$ & 5.95$\times10^{-4}$ \\
CO$_2$	& 9.10$\times10^{-18}$	& 1.68$\times10^{-5}$ & 5.81$\times10^{-5}$ \\
SiO	& 2.07$\times10^{-16}$	& 7.41$\times10^{-7}$ & 9.68$\times10^{-6}$ \\  \hline
\end{tabular}
\end{center}
\end{table}

To estimate the relevance of gas opacity for protoplanetary disks, we used an estimate of the gas ionization and photodissociation cross sections from \cite{gldh} to compute the mean photo dissociation cross section for various species in the wavelength range of CRUV photons (90-170 nm). Using this value, we computed the abundance that is required for an individual species to produce a gas opacity equal to the dust opacity. We did this for both molecular cloud and protoplanetary disk dust properties. \\

H$_2$ does not contribute to the opacity because most of it is in the fundamental vibrational level of the ground electronic state $X\, ^1\Sigma_g^+$. Therefore it is unable to absorb CRUV that are generated in transitions from excited electronic states to vibrationally excited levels of the $X\, ^1\Sigma_g^+$ state.\\

Using the results from Table \ref{tabgas}, we can check to which extent the simplifying assumption holds that gas opacity is negligible compared to that of the dust for CRUV photoprocesses.  At regions located at distances of r$\lesssim$8 AU chemical abundances of the listed molecules easily exceed the boundary values of Table \ref{tabgas}. This means that the effects of gas opacity have to be included in chemical models of the midplane of protoplanetary disks inside 10 AU.

\subsection{Chemistry at 1 AU}\label{r1}

\begin{figure}
\begin{center}
\includegraphics[scale=1.42]{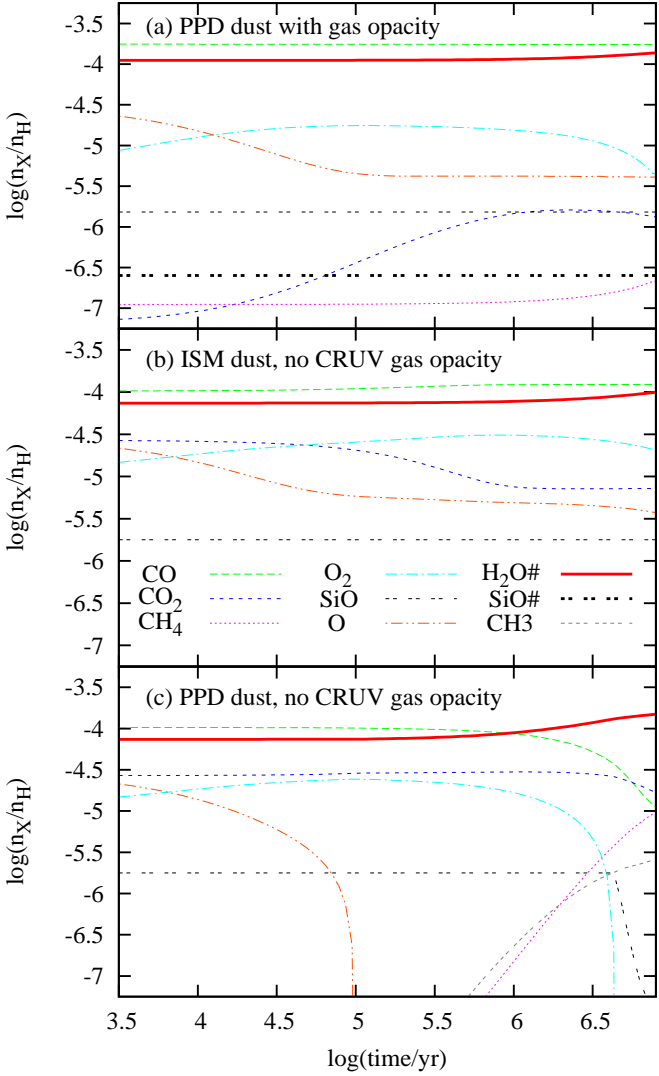}
\caption{Chemical abundance evolution at a distance of 1 AU from the central star: (a) Using an ISM dust value for  $\sigma_\mathrm{\langle H\rangle}^\mathrm{UV}$ and ignoring CRUV gas opacity, (b) using an appropriate $\sigma_\mathrm{\langle H\rangle}^\mathrm{UV}$ for protoplanetary disk conditions and ignoring CRUV gas opacity, and (c) using an appropriate $\sigma_\mathrm{\langle H\rangle}^\mathrm{UV}$ for protoplanetary disk conditions and including the effects of CRUV gas opacity.}\label{1au}
\end{center}
\end{figure}

In this region, the moderately high temperature (80 K) ensures that most of the chemical species stay in the gas phase. Adsorption of gas phase H$_2$O is much more efficient than thermal desorption, which causes gas phase H$_2$O to have a very low abundance. CO, O$_2$, CO$_2$, SiO, CH$_4$, and oxygen in atomic form are the dominant species in the gas phase, whereas H$_2$O and to a lesser extent SiO are the only species with a significant ice abundance. Fig. \ref{1au}a shows the time-dependent evolution of the species abundances when considering an appropriate protoplanetary disk dust size distribution\footnote{Our choice of dust size distribution parameters is typical of a Class II source, but limited to current knowledge based on observations.} and including the effects of CRUV gas opacity.\\

CO is very abundant in the initial conditions run, and remains the most abundant species besides H$_2$. CO is kept at this level through two main pathways. It can be photodissociated by CRUV into C and O, which later recombine in the following way:
\begin{equation}\label{cotoco}
  \mathrm{CO}\xrightarrow{\gamma_\mathrm{CRUV}}\mathrm{C,O;\ C\xrightarrow{H_2}CH_2\xrightarrow{O}CO}\ .
\end{equation}
CO is mainly formed via electron recombination of HCO$^+$. However, SiO also acts as a catalyst for the formation of CO because its reaction with HCO$^+$ is very rapid. Hence, whenever SiO is present, it efficiently converts HCO into CO, without SiO having to be very abundant itself. The OH to SiO to CO chemical pathway is the following:
\begin{equation}\label{ohtosio}
 \tx{OH\xr{O}O_2\xr{Si}SiO\xr{HCO^+}CO,SiOH^+;\ SiOH^+\xr{e-}Si,OH/SiO}\ .
\end{equation}  
HCO$^+$ is formed in the reaction of CO with H$_3^+$, which as noted above in Section \ref{crip} is a byproduct of cosmic-ray ionization of H$_2$. This cycle ensures a steady supply of OH as a primer for sustained formation of SiO and CO, as seen in Fig. \ref{1au}ab.\\
CH$_4$ is also steadily formed from He$^+$ ionization of CO:
\begin{equation}\label{cotoch4}
 \tx{CO\xr{He^+}C^+\xr{H_2}CH_{2\to3\to5}^+\xr{CO}CH_4}\ ,
\end{equation} 
although at a very low abundance (10$^{-7}$). The latter part of that reaction is curbed because the first step is reversed by O$_2$:
\begin{equation}\label{c+toco}
 \tx{C^++O_2\to CO+O^+} \ .
\end{equation} 
Atomic oxygen is depleted after 10$^4$ yr, favoring the formation of O$_2$ and CO$_2$. Abundances of CO$_2$ along with OH are enhanced due to the reactions
\begin{equation}\label{cotooh}
 \tx{CO\xr{H_3^+}HCO^+\xr{Fe}HCO\xr{O}OH,CO/CO_2}\ .
\end{equation} 
The OH enhancement is also curbed by the very efficient O$_2$/CO$_2$ formation reactions, which help keep the atomix oxygen abundance low:
\begin{equation}\label{ohtoo2}
 \tx{OH+O\to O_2+H}\ ,
\end{equation} 
\begin{equation}
 \tx{OH+CO\to CO_2+H}\ .
\end{equation} 
Atomic oxygen is constantly generated thanks to the steady CRUV photodissociation of CO, O$_2$, and CO$_2$. This atomic oxygen partly goes into the formation of water vapor, which is promptly adsorbed onto the grain surface due to the temperature and density conditions at 1 AU:
\begin{equation}\label{otoh2o}
 \tx{O\xr{H_3^+}OH^+,H_2O^+;\ OH^+\xr{H_2}H_2O^+\xr{H_2}H_3O^+\xr{e^-}H_2O/OH}\ .
\end{equation} 
This water vapor formation pathway explains the long-term ability of the system to keep forming water ice, through which it becomes the main oxygen reservoir at very long timescales. OH is recovered partly from SiOH$^+$ and also from the following H$_3$O$^+$ reaction, which can form water or OH at the same branching ratio:
\begin{equation}\label{h30+toh2o}
 \tx{H_3O^+\xr{SiO}H_2O,SiOH^+}\ .
\end{equation} 
\begin{figure}
\begin{center}
\includegraphics[angle=270, scale=0.7]{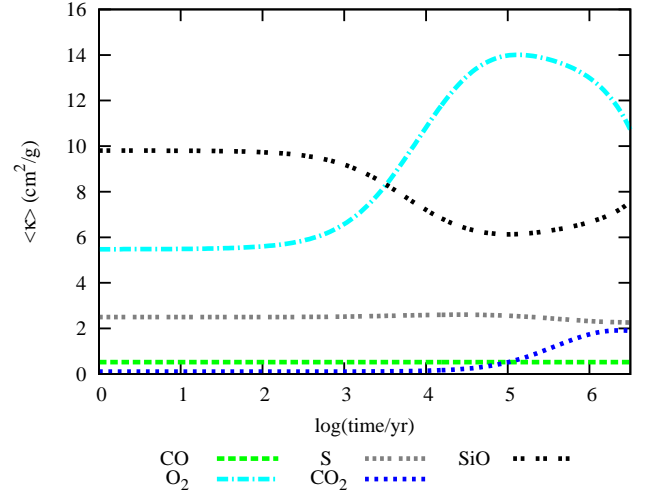}
\caption{CRUV gas opacity vs. time at 1 AU. For reference, the dust opacity is 68 cm$^2$/g (gas).}\label{opac1}
\end{center}
\end{figure}
Combining reactions (\ref{ohtosio}) and (\ref{h30+toh2o}) we can see how near the end, the oxygen in O$_2$ transfers to H$_2$O :
\begin{equation}\label{o2toh2o}
 \tx{O_2\xr{Si}SiO\xr{H_3O^+}H_2O}\ .
\end{equation} 
Gas-phase water created in the previous reaction freezes out almost immediately, as mentioned above (see Fig. \ref{1au}a). In Fig. \ref{opac1} we can see that after 10$^4$ yr the gas opacities reach a maximum of about 40\% with respect to the dust opacity, which remains constant\footnote{For simplicity, we did not include self-consistent grain growth models here.}. The late O$_2$ enhancement (Fig. \ref{1au}a) implies that it is providing most of the CRUV gas extinction, and consequently SiO absorbs fewer CRUV photons. Thus O$_2$ slows down atomic oxygen formation via photodissociation of SiO, which is a more efficient formation mechanism than the CRUV photodissociation of O$_2$. This causes a rebound effect after 10$^5$ yr, as O$_2$ needs a constant supply of atomic oxygen to form. Because OH is free to react with CO (instead of reacting with O), CO$_2$ formation is enhanced. However, CRUV photodissociated atomic oxygen from SiO and CO is necessary to sustain CO$_2$ production via reaction (\ref{cotooh}). This causes CO$_2$ to reach a saturation level near the end of the simulation. 

\begin{table}
 \begin{center}
\caption{Ratios between various species abundances and CO at 1 AU after 10$^6$ yr for three cases: Including the effects of gas opacity, considering only dust extinction from ISM dust parameters, and considering only dust extinction from protoplanetary disk dust parameters.}
\label{tabcomp}\renewcommand{\arraystretch}{1.4}
  \begin{tabular}{cccc}\hline\hline
   Ratio & With gas & \multicolumn{2}{c}{No gas opacity}\\ 
         & opacity, $\sigma_\mathrm{dust}^\mathrm{PPD}$ yr & $\sigma_\mathrm{dust}^\mathrm{ISM}$ & $\sigma_\mathrm{dust}^\mathrm{PPD}$ \\ \hline
    H$_2$O\#/CO & 0.69 & 1 & 0.66\\ 
    O$_2$/CO & 0.09 & 0.19 & 0.28\\
    CO$_2$/CO & 0.01 & 0.33 & 0.07\\
    CH$_4$/CO & 7$\times$10$^{-4}$ & 2$\times$10$^{-3}$ & 10$^{-5}$\\ \hline
  \end{tabular}
 \end{center}
\end{table}

\subsubsection{Comparison to our previous work}

To understand the effects of gas opacity, we also ran the simulation considering dust as the only sink for CRUV photons. We used the simplified formula for CRUV photoprocess rates from \cite{paper1}, where the total extinction cross section is simply the grain extinction cross section. Since this calculation is not dependent on the gas phase abundances, the rate coefficients are constant throughout the simulation.\\

When considering a UV extinction cross section based on a protoplanetary disk dust size distribution, the CRUV photon flux is higher than for ISM dust parameters \citep{paper1}. This is because dust coagulation processes in protoplanetary disks increase the average size of dust grains, which in turn decreases the effective area for dust absorption of UV photons. For protoplanetary disk conditions the UV extinction cross section $\sigma_\mathrm{\langle H\rangle}^\mathrm{UV}$ is 13.5 times lower than the typical ISM value, $\sigma_\mathrm{\langle H\rangle}^\mathrm{UV}=2\times10^{-21}$ cm$^{-1}$.\\

Fig. \ref{1au}b shows the time-dependent evolution of the chemical abundances for the ISM-like dust parameters used adopted for the calculation of CRUV rates in \textsc{Umist}. Here, the main carriers of carbon are CO and CO$_2$, followed by the slightly less abundant CH$_4$ and CH$_3$. After a few 10$^5$ years the carbon is almost equally divided among CH$_4$, CO, and CO$_2$.  SiO is the second most important gas phase oxygen carrier up to a few Myr, when it stops to form. This is because the grains are cool enough to hold a large amount of water ice on their surface, effectively stopping OH formation and the SiO formation pathway (\ref{ohtosio}). As the oxygen in CO, CO$_2$, and SiO is transfered to water ice near the end of the disk lifetime, carbon binds with H$_3^+$ and H$_2$, which enhances the formation rates of molecules such as CH$_3$ and CH$_4$.\\


\begin{figure}
\begin{center}
\includegraphics[angle=270, scale=0.7]{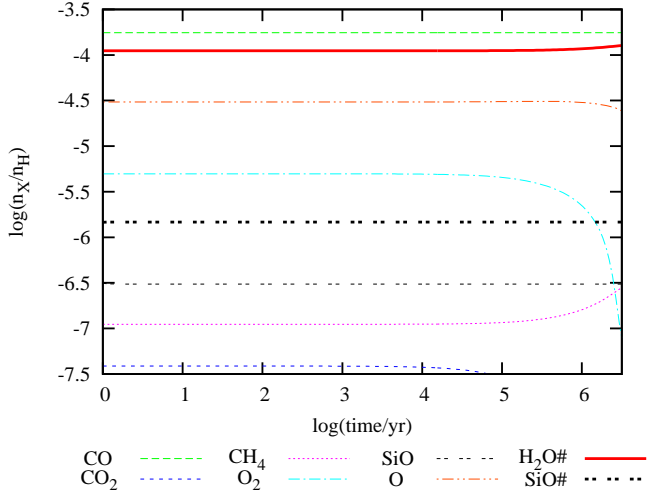}
\caption{Chemical abundance evolution at a distance of 3-5 AU from the central star.}\label{4au}
\end{center}
\end{figure}

When comparing the chemical evolution obtained when using ISM dust parameters (Fig. \ref{1au}b) vs. appropriate protoplanetary disk dust parameters (Fig. \ref{1au}c), the impact of the larger CRUV flux on the abundance of various species is evidenced. As mentioned above, CRUV photodissociation processes are up to 13.5 times more efficient in protoplanetary disks than when estimated using the interstellar UV extinction cross section value used in \textsc{Umist} \citep{paper1}. Because the effects of gas extinction are neglected, the CRUV flux is enhanced uniformly for all species. This affects specifically the long-term formation of SiO, O$_2$, and hydrocarbons. For instance, CH$_3$ and CH$_4$ are not efficiently formed because they are very sensitive to CRUV photoprocesses. CH$_4$ photodissociates into CH$_2$, and CH$_3$ photodissociates into CH, CH$_2$ and CH$_3^+$. If the CRUV field is strong enough, CH$_3$ and CH$_4$ will be destroyed at the expense of CO on a timescale similar to their formation timescale.\\

The CRUV photodissociation of CO, SiO, and CO$_2$ is very efficient for high CRUV fluxes. These photoprocesses steadily produce atomic oxygen at a higher rate than when considering the effects of gas opacity. This efficient formation of atomic oxygen forms OH from HCO in reaction (\ref{cotooh}) and preserves the OH, SiO cycle in reaction (\ref{ohtosio}). This means that SiO is never depleted even at extremely long timescales (below 10$^8$ yr). Even though CO$_2$ is also created from OH at the expense of CO, it never reaches a high abundance because it is destroyed via CRUV photodissociation, which ensures that the CO$_2$ abundances stay low compared to those seen in Fig. \ref{1au}a.\\

Most notably, the CO$_2$/CO, O$_2$/CO, and CO/CH$_{3,4}$ ratios change enormously between the three cases, as shown in Table \ref{tabcomp}. This shows that the chemical evolution obtained when considering the effects of gas opacity cannot be obtained by a simple interpolation between a low and a high CRUV field, but it has to be studied for each species separately.

\subsection{Chemistry at 3-5 AU}
\begin{figure}
\begin{center}
\includegraphics[angle=270, scale=0.7]{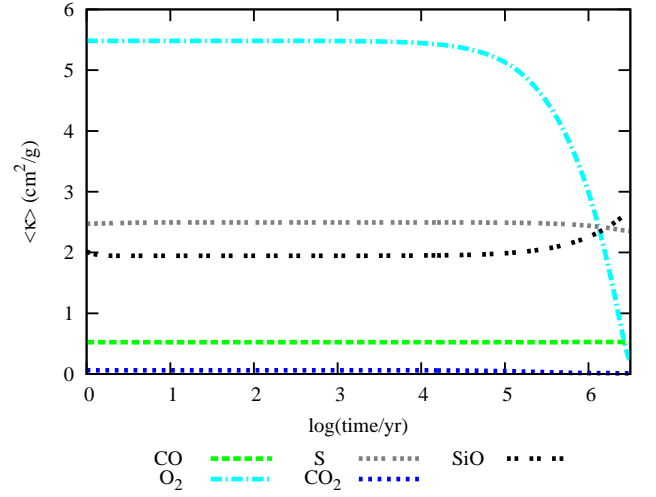}
\caption{CRUV gas opacity vs. time at a distance of 3-5 AU. For reference, the dust opacity is 68 cm$^2$/g (gas).}\label{opac4}
\end{center}
\end{figure}
Figure \ref{4au} shows the chemical evolution in a region located at 3-5 AU from the central star, where $n_\tx{\langle H\rangle}$=10$^{12}$ cm$^{-3}$ and $T$=65 K \citep{wkt}. In this region, reactions (\ref{cotoco}) ,(\ref{cotoch4}), (\ref{o2toh2o}), and (\ref{otoh2o}) still take place. However, as the temperature here is lower than at 1 AU, SiO is now mostly in the ice phase, and retains a small gas phase abundance. It is kept at a stable but low level throughout the simulation by the following feedback reaction, similar to reaction (\ref{ohtosio}):
\begin{equation}
 \tx{SiO\xr{H^+}SiO^+\xr{H_2}SiOH^+\xr{e^-}SiO/Si,OH}\ .
\end{equation} 
Therfore, OH formation is not as efficient as it was at 1 AU. The low OH formation rate cannot sustain the formation of O$_2$ in reaction (\ref{ohtoo2}) when it is forming H$_2$O, and the oxygen in O$_2$ transfers to H$_2$O via reaction \ref{o2toh2o}. As the O$_2$ abundance starts to decrease, the backwards reaction (\ref{c+toco}) is stopped, which allows CO to form CH$_4$ more efficiently via reaction \ref{cotoch4} (see Fig. \ref{4au}). The consequent decrease in O$_2$ opacity seen in Fig. \ref{opac4} causes the CRUV photorates to increase for other molecules, as it allows more CRUV photons to dissociate or ionize other species. This is shown in Fig \ref{opac4}, where the SiO opacity increases as the O$_2$ opacity decreases. Now CH$_4$ CRUV photodissociation becomes very efficient at forming CH$_2$, which causes the following reaction 
\begin{equation}
 \tx{CH_2+O\to CO+H+H}\ ,
\end{equation} 
to be favored over reaction (\ref{ohtoo2}). This exacerbates the depletion of O$_2$, as seen at the end of the simulation in Fig. \ref{4au}. The gas opacity in this region (Fig. \ref{opac4}) is less than half the value at 1 AU (Fig. \ref{opac1}, note the different vertical scale), which is caused by the lack of efficient O$_2$ formation and the freeze-out of SiO. The first part of reaction (\ref{cotooh}) is reversed by the following reaction:
\begin{equation}\label{hco+toco}
 \tx{HCO^++e^-\to CO+H}\ ,
\end{equation} 
which keeps the CO$_2$ levels low. This is enabled by an increase in effectivity of electron recombination reactions vs. neutral reactions brought about by the low-density conditions, compared to the 1 AU region.

\subsection{Chemistry at 7-8 AU}

\begin{figure}
\begin{center}
\includegraphics[angle=270, scale=0.7]{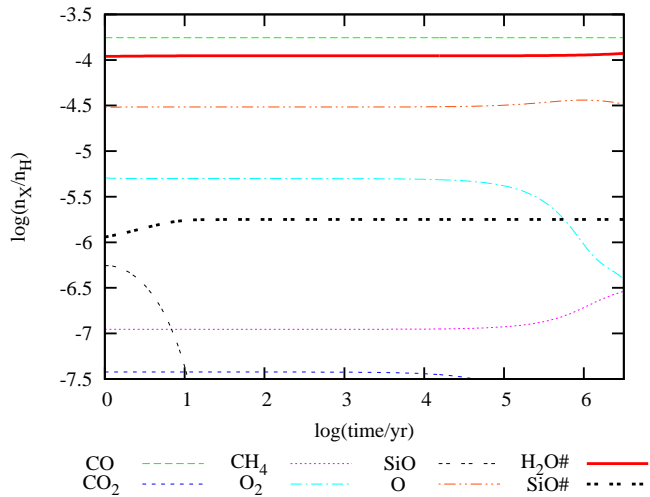}
\caption{Chemical abundance evolution at a distance of 7-8 AU from the central star.}\label{8au}
\end{center}
\end{figure}

In this region, the physical conditions are $n_\tx{\langle H\rangle}$=10$^{11}$ cm$^{-3}$ and $T$=50 K \citep{wkt}. As we can see from the chemical evolution plot in Fig. \ref{8au}, H$_2$O and SiO are completely frozen after only 1 yr, with negligible gas phase abundances. However, the chemical evolution for the other significant species remains more or less the same. The lack of SiO in the gas phase impacts the late formation of H$_2$O that appears at 1-5 AU. Since O$_2$ is not consumed via reaction (\ref{o2toh2o}, its depletion is not as dramatic as in the 3-5 AU region. Now the main reaction that removes O$_2$ from the gas is
\begin{equation}
 \tx{C+O_2\to CO+O}\ .
\end{equation} 
This reaction, which does not have an activation energy barrier and is very rapid at very low temperatures \citep{smith}, allows reaction (\ref{cotoch4}) to form CH$_4$ very late in the simulation in the same way that we observed for the previous region (3-5 AU). Reactions (\ref{cotoco},\ref{ohtoo2},\ref{otoh2o}) are also responsible for the formation of CO, O$_2$, and H$_2$O, respectively. Just as in the 3-5 AU region, CO$_2$ is not efficiently formed because its formation via reaction (\ref{cotooh}) is stopped by the backreaction (\ref{hco+toco}). The gas opacity here (Fig. \ref{opac8}) is very similar to that observed at 3-5 AU (Fig. \ref{opac4}). The absence of SiO opacity due to freeze-out enhances the CRUV rates for all other species, including CO.

\section{Discussion}\label{disc}
We have included the gas opacity of all atoms and molecules that are photoprocessed by cosmic-ray-induced UV photons in our chemical evolution model. By doing this, we were able to assess whether a given species is relevant for the extinction of CRUV in the midplane of a protoplanetary disks. By studying the impact that the gas opacity of individual species has on the chemistry and vice versa, we can simplify the chemical model by including only the most important species. As some opacities increase, an important shielding effect appears and changes the CRUV photoprocess rates. This affects the chemical formation pathways for both gas- and ice-phase species. \\
\begin{figure}
\begin{center}
\includegraphics[angle=270, scale=0.7]{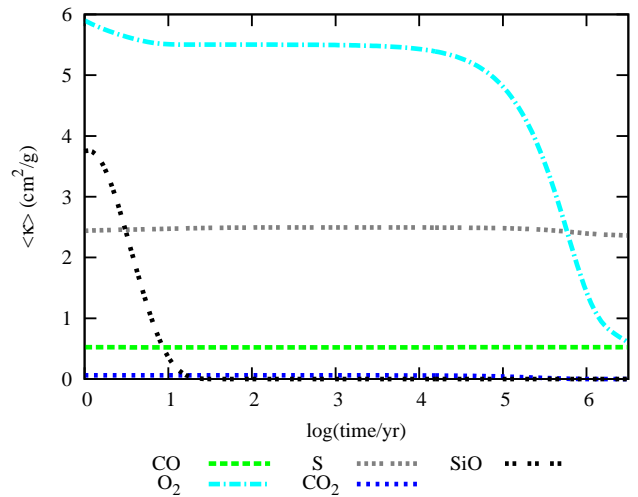}
\caption{CRUV gas opacity vs. time at a distance of 7-8 AU. For reference, the dust opacity is 68 cm$^2$/g (gas).}\label{opac8}
\end{center}
\end{figure}

As discussed in Section \ref{cm}, the most important species that contribute to the gas opacity are O$_2$, SiO, S, CO, and CO$_2$ consistently throughout the 1-10 AU region of the disk midplane. We found that the opacities of O$_2$, SiO (and CO$_2$ in the warm regions around 1 AU) are the most variable in time and hence are the ones that require most attention. As the CO abundance remains more or less constant in those regions, its opacity remains fairly constant as well. Even though we did not include sulphur chemistry beyond ionization, its CRUV opacity ranges from 2.5 to 3.5 cm$^2$/g gas (see Figs. \ref{opac1}, \ref{opac4}, \ref{opac8} and \ref{opac10}), reaching its highest contribution with respect to the total gas opacity (70\%) and with respect to the total dust and gas opacity (3.4\%) at 10-15 AU. \\

When comparing the gas opacity to the total CRUV opacity in Table \ref{tabopac}, it is evident that its effect cannot be easily dismissed. In the region 1-2 AU the contribution from the gas is up to 30\% of the combined gas and dust opacity. At 7-8 AU, the contribution is up to 10\%, mostly because of SiO freeze-out. Finally, at 10 AU the temperature becomes sufficiently low to freeze CO, CO$_2$ and CH$_4$ onto the grain surface, and the contribution of the gas opacity becomes very low (5\%). Therefore, future chemical models that aim to include this treatment of CRUV gas and dust opacity can consider the contribution of the most significant species that our models yield for the probed regions of the disk.\\

The general chemical composition of the disk midplane obtained using our models compare favorably to that in \cite{walsh}, particularly in the freeze-out of water at distances above 1 AU, and in the significant gas phase presence of CO (and CO$_2$ to a lesser degree) throughout the studied region. We notice the same low abundances of H$_2$CO and OH. These similarities are caused by the temperature structure, which controls the adsorption and desorption processes.

\subsection{The role of OH}

The availability of OH caused by the high and steady abundance of SiO at 1 AU causes a late O and O$_2$ enhancement.  The products of SiO dissociation later form H$_3$O$^+$, which via dissociative recombinations (Si, e-) forms water vapor. Atomic oxygen is very abundant at 1 AU thanks to cosmic-ray-induced dissociation of CO and SiO. However, oxygen-driven OH formation acts as a catalyst for steady CO formation. A rebound effect is observed at a late stage (10$^5$ yr), as O$_2$ self-shields the CRUV photons that drive the OH driven pathways for its formation.\\

\begin{figure}
\begin{center}
\includegraphics[angle=270, scale=0.7]{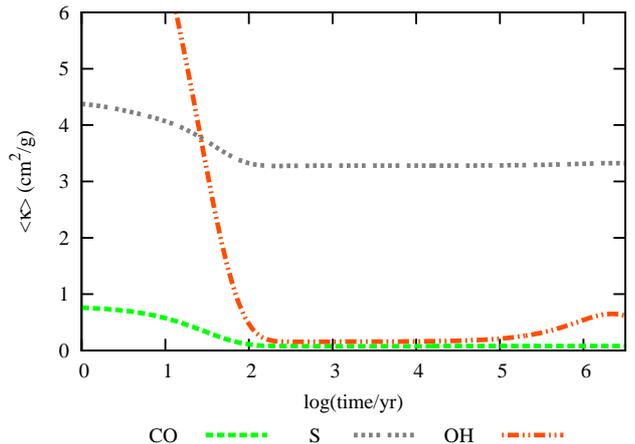}
\caption{CRUV gas opacity vs. time at a distance of 10 AU. For reference, the dust opacity is 68 cm$^2$/g gas).}\label{opac10}
\end{center}
\end{figure}

Even though SiO is formed in reaction (\ref{ohtosio}) at 3-5 AU, under these conditions it is very efficiently adsorbed onto the grain surface. This affects the formation of OH and thus O$_2$ is depleted at long timescales, favoring CO formation via CH$_4$ and later CH$_2$ reactions with O. As the gas phase SiO is depleted, no O$_2$ is transformed into H$_2$O, which avoids the complete depletion of O$_2$. At 7-8 AU there is an overall enhancement of CRUV rates because of this. After 10 AU, the total gas opacity is sufficiently low for OH to drive the chemistry. Indeed, Fig. \ref{opac10} shows the gas opacity at 10 AU, in which OH is seen to contribute. This agrees with the results from our previous research paper \citep{paper1}.

\begin{table}
 \begin{center}
\caption{Total gas opacity as a function of time and distance. Percentage relative to the total dust (68 cm$^2$/g gas) and gas opacity is given in parentheses.}
\label{tabopac}\renewcommand{\arraystretch}{1.4}
  \begin{tabular}{cccc}\hline\hline
   Distance &  \multicolumn{3}{c}{Total gas opacity $\langle\kappa_\mathrm{gas}\rangle$ (cm$^2$/g gas)}\\ 
         & 10$^4$ yr & 10$^5$ yr & 10$^6$ yr \\ \hline
    1 AU & 23.6 (25.8\%) & 25.4 (27.2\%) & 26.7 (28.2\%)\\ 
    3-5 AU & 11.5 (14.5\%)& 11.1 (14\%)& 9.2 (12\%)\\
    7-8 AU & 9.5 (12.3\%)& 9 (11.7\%)& 5.4 (7.4\%)\\
    10-15 AU& 3.6 (5\%)& 3.7 (5.2\%)& 4.2 (5.8\%)\\ \hline
  \end{tabular}
 \end{center}
\end{table}
\subsection{Survival of SiO in the $A_V=1$ region}\label{disc2}

The longevity of SiO in our models of the disk midplane (see Figs. \ref{1au}a and \ref{4au}) could have potential consequences beyond the midplane of the disk. We explored the chemistry at 1 AU in a direction perpendicular to the disk plane upwards to densities of $n_\mathrm{\langle H\rangle}=10^{11}$ cm$^{-3}$. We saw that the abundance of SiO remains steadily high, meaning that SiO can be formed even above the midplane.\\

Following \cite{aikawadif}, the vertical drift timescale $\tau_\mathrm{vd}$ of a molecule formed in the midplane of the disk that moves up to a distance $z=\lambda$ above the midplane can be estimated via the disk viscosity\footnote{The value of $\nu=\alpha c_s h$ depends on the sound speed $c_s$, the scale height $h$ of the disk and the Shakura Sunyaev $\alpha$ parameter, for which we chose a value of 0.1 corresponding to a young disk with $\dot{M}\simeq10^{-8}M_\odot/yr$. This mass accretion rate value suggests an age for the disk in the range 10$^{5.5}$-10$^{6.5}$ yr, according to \cite{hartmann}.} $\nu$ \citep{visc}:
\begin{equation}
 \tau_\mathrm{vd}=\frac{\lambda^2}{\nu}\ .
\end{equation} 
A simple calculation for $\lambda=0.1$ AU, corresponding to the $A_V=1$ region (at $r=1$ AU) of the disk, yields
\begin{equation}
 \tau_\mathrm{vd}\simeq280\ \mathrm{yr}\ .
\end{equation} 
If the SiO photodissociation timescale at $A_V=1$ is longer than the vertical drift timescale, it should be possible for SiO to accumulate on the surface where it could be observed. We can roughly estimate this timescale, which depends on the strength of the local UV field $\chi$. Using the detailed radiative transfer model of \cite{wkt}, we derive a value for $\chi$ in the range 1-10 at the $A_V=1$ line. Accordingly, an SiO molecule can survive for approximately 10$^3$ yr if $\chi=1$. This means that it may be possible for SiO to accumulate around $A_V=1$ and be detectable. For example, SiO masers are detected in the circumstellar regions around asymptotic giant branch (AGB) stars, and their maser lines are found to originate from regions below the dust evaporation radius \citep{agb}. More recently, gas phase SiO has been detected in the debris disk around $\eta$ Corvi \citep{etacorvi}. Hence it should be possible to detect accumulated SiO even if there are dust grains present. Our low ISM metal abundances already take Si incorporation into dust grains into account. Given that the region where SiO is being formed has a small radial extent ($r\le1-5$ AU), detection of SiO could prove to be very difficult in the near future, except for high-sensitivity, high-spatial-resolution observations that ALMA could carry out.

\section{Conclusions}\label{conc}

We have developed a tool for including the effects of gas extinction of cosmic-ray-induced UV photons in the chemical evolution in the disk midplane. The most important species that contribute to the gas opacity in the 1-8 AU region of the disk midplane are O$_2$, SiO, S, CO, and CO$_2$. Because the CO abundance is steady in those regions, its opacity remains fairly constant as well. Even though CO competes with water ice for being the most abundant species, CO is not a major contributor to the opacity, whereas CO$_2$ play a more important role in the opacity while not being as chemically abundant as CO.\\

We find that the opacities of O$_2$, SiO (and CO$_2$ in the warm regions around 1 AU) are highly variable in time, and their effects on the gas and ice chemistry cannot be easily dismissed. Therefore, future models should include the gas opacity of these species in the calculation of CRUV photoprocess rates.\\

As most of the emission in recent IR observations of OH and H$_2$O in disks \citep{carr1,carr2,fedele2,meeus} is likely coming from layers well above the midplane, the catalytic role of OH in the midplane is not necessarily something we can prove by observing its surface abundance. \\

Finally, we found that if SiO is steadily created in the midplane at long timescales, it can drift vertically upwards and accumulate around the $A_V=1$ region, which suggests a possibility for future detections of SiO in protoplanetary disks.

\appendix
\section{Cross section data}

\begin{table}[ht]
\begin{center}
\caption{Cosmic ray-induced UV photoprocess list.}
\label{tabproc}
\begin{tabular}{llccc}\hline\hline
 Species & Proc. & Products & Source & Type \\ \hline
 C & Ion. & C$^+$+e- & Leiden & Cont. \\ 
 CH & Dis. & C+H & Leiden & Line/Cont.\\ 
 CH$^+$ & Dis. & C$^+$+H & Leiden & Line/Cont. \\ 
 CH$_2$ & Ion. & CH$_2^+$+e- & UMIST06 & Cont.\\
 & & & \citep{gldh} & \\ 
 CH$_2$ & Dis. & CH+H & Leiden & Line/Cont. \\ 
 CH$_3$ & Ion. & CH$_3^+$+e- & UMIST06 & Cont.\\
 & & & (No ref. data) & \\
 CH$_3$ & Dis. & CH$_2$+H & Leiden & Line \\ 
 CH$_3$ & Dis. & CH+H$_2$ & Leiden & Line \\ 
 CH$_4$ & Dis. & CH$_2$+H$_2$ & Leiden & Line/Cont. \\ 
 OH & Dis. & O+H & Leiden & Line/Cont. \\ 
 H$_2$O & Dis. & OH+H & Leiden & Cont. \\ 
 Mg & Ion. & Mg$^+$+e- & Leiden & Line/Cont. \\ 
 CO & Dis. & C+O & Leiden & Line \\ 
 Si & Ion. & Si$^+$+e- & Leiden & Cont. \\ 
 HCO & Dis. & CO+H & Leiden & Line \\ 
 HCO & Ion. & HCO$^+$+e- & UMIST06 & Cont. \\ 
 & & & \citep{gldh} & \\
 SiH & Dis. &  Si+H  & Leiden & Line \\ 
 H$_2$CO & Dis. & CO+H$_2$ & Leiden & Line/Cont. \\ 
 O$_2$ & Dis. & O+O & Leiden & Line/Cont. \\ 
 O$_2$ & Ion. & O$_2^+$+e- & Leiden & Line \\ 
 S & Ion. & S$^+$+e- & Leiden & Cont. \\ 
 CO$_2$ & Dis. & CO+O  & Leiden & Line/Cont. \\ 
 SiO & Dis. & Si+O  & Leiden & Line \\ 
 Fe & Ion. & Fe$^+$+e-  & Leiden & Cont. \\ \hline
\end{tabular}
\end{center}
\end{table}
Table \ref{tabproc} presents the information on all CRUV photoprocesses considered here, based on the UMIST06 reaction database \citep{umist}. Column 2 lists whether the photoprocess ionizes or dissociates the species. Column 3 shows the products of the reaction. Column 4 references the source for the photoprocess cross section, which for most of them is Ewine van Dishoeck's Leiden photoprocess database \texttt{http://www.strw.leidenuniv.nl/~ewine/photo/} \citep{faraday}. Column 5 refers to whether the photoprocess is a discrete and/or continuum absorption process.
\begin{acknowledgements}
We would like to thank Malcolm Walmsley for asking the basic question that triggered this research.
\end{acknowledgements}

\bibliographystyle{aa}
\bibliography{bib}

\begin{thebibliography}{54}
\expandafter\ifx\csname natexlab\endcsname\relax\def\natexlab#1{#1}\fi

\bibitem[{{Abgrall} {et~al.}(2000){Abgrall}, {Roueff}, \& {Drira}}]{abgrall}
{Abgrall}, H., {Roueff}, E., \& {Drira}, I. 2000, \aaps, 141, 297

\bibitem[{{Aikawa}(2007)}]{aikawadif}
{Aikawa}, Y. 2007, \apjl, 656, L93

\bibitem[{{Aikawa} {et~al.}(1996){Aikawa}, {Miyama}, {Nakano}, \&
  {Umebayashi}}]{aikawa}
{Aikawa}, Y., {Miyama}, S.~M., {Nakano}, T., \& {Umebayashi}, T. 1996, \apj,
  467, 684

\bibitem[{{Aresu} {et~al.}(2011){Aresu}, {Kamp}, {Meijerink}, {Woitke}, {Thi},
  \& {Spaans}}]{giamba1}
{Aresu}, G., {Kamp}, I., {Meijerink}, R., {et~al.} 2011, \aap, 526, A163

\bibitem[{Brown {et~al.}(1989)Brown, Byrne, \& Hindmarsh}]{vode}
Brown, P.~N., Byrne, G.~D., \& Hindmarsh, A.~C. 1989, SIAM J. Sci. Stat.
  Comput., 10, 1038

\bibitem[{{Carr} \& {Najita}(2008)}]{carr1}
{Carr}, J.~S. \& {Najita}, J.~R. 2008, Science, 319, 1504

\bibitem[{{Cazaux} \& {Tielens}(2002)}]{cazaux}
{Cazaux}, S. \& {Tielens}, A.~G.~G.~M. 2002, \apjl, 575, L29

\bibitem[{{Cecchi-Pestellini} \& {Aiello}(1992)}]{cecchi}
{Cecchi-Pestellini}, C. \& {Aiello}, S. 1992, \mnras, 258, 125

\bibitem[{{Chaparro Molano} \& {Kamp}(2012)}]{paper1}
{Chaparro Molano}, G. \& {Kamp}, I. 2012, \aap, 537, A138

\bibitem[{{D'Alessio} {et~al.}(2001){D'Alessio}, {Calvet}, \&
  {Hartmann}}]{dalessio}
{D'Alessio}, P., {Calvet}, N., \& {Hartmann}, L. 2001, \apj, 553, 321

\bibitem[{{Fedele} {et~al.}(2011){Fedele}, {Pascucci}, {Brittain}, {Kamp},
  {Woitke}, {Williams}, {Dent}, \& {Thi}}]{fedele2}
{Fedele}, D., {Pascucci}, I., {Brittain}, S., {et~al.} 2011, \apj, 732, 106

\bibitem[{{Fedele} {et~al.}(2010){Fedele}, {van den Ancker}, {Henning},
  {Jayawardhana}, \& {Oliveira}}]{fedele}
{Fedele}, D., {van den Ancker}, M.~E., {Henning}, T., {Jayawardhana}, R., \&
  {Oliveira}, J.~M. 2010, \aap, 510, A72

\bibitem[{{Glassgold} {et~al.}(2007){Glassgold}, {Najita}, \&
  {Igea}}]{glassgold}
{Glassgold}, A.~E., {Najita}, J.~R., \& {Igea}, J. 2007, \apj, 656, 515

\bibitem[{{Gorti} {et~al.}(2011){Gorti}, {Hollenbach}, {Najita}, \&
  {Pascucci}}]{gorti}
{Gorti}, U., {Hollenbach}, D., {Najita}, J., \& {Pascucci}, I. 2011, \apj, 735,
  90

\bibitem[{{Graedel} {et~al.}(1982){Graedel}, {Langer}, \& {Frerking}}]{graedel}
{Graedel}, T.~E., {Langer}, W.~D., \& {Frerking}, M.~A. 1982, \apjs, 48, 321

\bibitem[{{Gredel} {et~al.}(1987){Gredel}, {Lepp}, \& {Dalgarno}}]{gld}
{Gredel}, R., {Lepp}, S., \& {Dalgarno}, A. 1987, \apjl, 323, L137

\bibitem[{{Gredel} {et~al.}(1989){Gredel}, {Lepp}, {Dalgarno}, \&
  {Herbst}}]{gldh}
{Gredel}, R., {Lepp}, S., {Dalgarno}, A., \& {Herbst}, E. 1989, \apj, 347, 289

\bibitem[{{Habing}(1996)}]{agb}
{Habing}, H.~J. 1996, \aapr, 7, 97

\bibitem[{{Haisch} {et~al.}(2010){Haisch}, {Barsony}, \& {Tinney}}]{haisch}
{Haisch}, Jr., K.~E., {Barsony}, M., \& {Tinney}, C. 2010, \apjl, 719, L90

\bibitem[{{Hartmann} {et~al.}(1998){Hartmann}, {Calvet}, {Gullbring}, \&
  {D'Alessio}}]{hartmann}
{Hartmann}, L., {Calvet}, N., {Gullbring}, E., \& {D'Alessio}, P. 1998, \apj,
  495, 385

\bibitem[{{Hasegawa} \& {Herbst}(1993)}]{herbst}
{Hasegawa}, T.~I. \& {Herbst}, E. 1993, \mnras, 261, 83

\bibitem[{{Herbst} \& {Klemperer}(1973)}]{klemperer}
{Herbst}, E. \& {Klemperer}, W. 1973, \apj, 185, 505

\bibitem[{{Indriolo} {et~al.}(2007){Indriolo}, {Geballe}, {Oka}, \&
  {McCall}}]{indriolo}
{Indriolo}, N., {Geballe}, T.~R., {Oka}, T., \& {McCall}, B.~J. 2007, \apj,
  671, 1736

\bibitem[{{Jenkins}(2009)}]{jenkins}
{Jenkins}, E.~B. 2009, \apj, 700, 1299

\bibitem[{{Lee} {et~al.}(1996){Lee}, {Bettens}, \& {Herbst}}]{lee2}
{Lee}, H.-H., {Bettens}, R.~P.~A., \& {Herbst}, E. 1996, \aaps, 119, 111

\bibitem[{{Leger} {et~al.}(1985){Leger}, {Jura}, \& {Omont}}]{leger}
{Leger}, A., {Jura}, M., \& {Omont}, A. 1985, \aap, 144, 147

\bibitem[{{Lisse} {et~al.}(2012){Lisse}, {Wyatt}, {Chen}, {Morlok}, {Watson},
  {Manoj}, {Sheehan}, {Currie}, {Thebault}, \& {Sitko}}]{etacorvi}
{Lisse}, C.~M., {Wyatt}, M.~C., {Chen}, C.~H., {et~al.} 2012, \apj, 747, 93

\bibitem[{{Meeus} {et~al.}(2012){Meeus}, {Montesinos}, {Mendigut{\'{\i}}a},
  {Kamp}, {Thi}, {Eiroa}, {Grady}, {Mathews}, {Sandell}, {Martin-Za{\"i}di},
  {Brittain}, {Dent}, {Howard}, {M{\'e}nard}, {Pinte}, {Roberge},
  {Vandenbussche}, \& {Williams}}]{meeus}
{Meeus}, G., {Montesinos}, B., {Mendigut{\'{\i}}a}, I., {et~al.} 2012, \aap,
  544, A78

\bibitem[{{Micelotta} {et~al.}(2011){Micelotta}, {Jones}, \& {Tielens}}]{micel}
{Micelotta}, E.~R., {Jones}, A.~P., \& {Tielens}, A.~G.~G.~M. 2011, \aap, 526,
  A52

\bibitem[{{Najita} {et~al.}(2010){Najita}, {Carr}, {Strom}, {Watson},
  {Pascucci}, {Hollenbach}, {Gorti}, \& {Keller}}]{carr2}
{Najita}, J.~R., {Carr}, J.~S., {Strom}, S.~E., {et~al.} 2010, \apj, 712, 274

\bibitem[{{{\"O}berg} {et~al.}(2009){{\"O}berg}, {van Dishoeck}, \&
  {Linnartz}}]{oberg}
{{\"O}berg}, K.~I., {van Dishoeck}, E.~F., \& {Linnartz}, H. 2009, \aap, 496,
  281

\bibitem[{Ochkin \& Kittell(2009)}]{plasma}
Ochkin, V. \& Kittell, S. 2009, Spectroscopy of low temperature plasma
  (Wiley-VCH)

\bibitem[{{Padovani} \& {Galli}(2011)}]{padov}
{Padovani}, M. \& {Galli}, D. 2011, \aap, 530, A109

\bibitem[{{Prasad} \& {Tarafdar}(1983)}]{prasad}
{Prasad}, S.~S. \& {Tarafdar}, S.~P. 1983, \apj, 267, 603

\bibitem[{{Riahi} {et~al.}(2006){Riahi}, {Teulet}, {Ben Lakhdar}, \&
  {Gleizes}}]{riahi}
{Riahi}, R., {Teulet}, P., {Ben Lakhdar}, Z., \& {Gleizes}, A. 2006, Eur. Phys.
  J. D, 40, 223

\bibitem[{{Roberts} {et~al.}(2007){Roberts}, {Rawlings}, {Viti}, \&
  {Williams}}]{roberts}
{Roberts}, J.~F., {Rawlings}, J.~M.~C., {Viti}, S., \& {Williams}, D.~A. 2007,
  \mnras, 382, 733

\bibitem[{{Semenov} {et~al.}(2010){Semenov}, {Hersant}, {Wakelam}, {Dutrey},
  {Chapillon}, {Guilloteau}, {Henning}, {Launhardt}, {Pi{\'e}tu}, \&
  {Schreyer}}]{semenov}
{Semenov}, D., {Hersant}, F., {Wakelam}, V., {et~al.} 2010, \aap, 522, A42

\bibitem[{{Semenov} \& {Wiebe}(2011)}]{wiebe}
{Semenov}, D. \& {Wiebe}, D. 2011, \apjs, 196, 25

\bibitem[{{Semenov} {et~al.}(2004){Semenov}, {Wiebe}, \& {Henning}}]{semcr}
{Semenov}, D., {Wiebe}, D., \& {Henning}, T. 2004, \aap, 417, 93

\bibitem[{{Shakura} \& {Sunyaev}(1973)}]{visc}
{Shakura}, N.~I. \& {Sunyaev}, R.~A. 1973, \aap, 24, 337

\bibitem[{{Shen} {et~al.}(2004){Shen}, {Greenberg}, {Schutte}, \& {van
  Dishoeck}}]{shenvd}
{Shen}, C.~J., {Greenberg}, J.~M., {Schutte}, W.~A., \& {van Dishoeck}, E.~F.
  2004, \aap, 415, 203

\bibitem[{{Smith} {et~al.}(2004){Smith}, {Herbst}, \& {Chang}}]{smith}
{Smith}, I.~W.~M., {Herbst}, E., \& {Chang}, Q. 2004, \mnras, 350, 323

\bibitem[{{Sternberg} \& {Dalgarno}(1995)}]{sd}
{Sternberg}, A. \& {Dalgarno}, A. 1995, \apjs, 99, 565

\bibitem[{{Sternberg} {et~al.}(1987){Sternberg}, {Dalgarno}, \& {Lepp}}]{sdl}
{Sternberg}, A., {Dalgarno}, A., \& {Lepp}, S. 1987, \apj, 320, 676

\bibitem[{{Thi} {et~al.}(2011){Thi}, {Woitke}, \& {Kamp}}]{thi}
{Thi}, W.-F., {Woitke}, P., \& {Kamp}, I. 2011, \mnras, 412, 711

\bibitem[{{Umebayashi} \& {Nakano}(1981)}]{umebayashi}
{Umebayashi}, T. \& {Nakano}, T. 1981, \pasj, 33, 617

\bibitem[{{van Dishoeck} {et~al.}(2006){van Dishoeck}, {Jonkheid}, \& {van
  Hemert}}]{faraday}
{van Dishoeck}, E.~F., {Jonkheid}, B., \& {van Hemert}, M.~C. 2006, Faraday
  Discussions, 133, 231

\bibitem[{{van Zadelhoff} {et~al.}(2001){van Zadelhoff}, {van Dishoeck}, {Thi},
  \& {Blake}}]{zadel}
{van Zadelhoff}, G.-J., {van Dishoeck}, E.~F., {Thi}, W.-F., \& {Blake}, G.~A.
  2001, \aap, 377, 566

\bibitem[{{Visser} {et~al.}(2011){Visser}, {Doty}, \& {van Dishoeck}}]{visser2}
{Visser}, R., {Doty}, S.~D., \& {van Dishoeck}, E.~F. 2011, \aap, 534, A132

\bibitem[{{Visser} {et~al.}(2009){Visser}, {van Dishoeck}, {Doty}, \&
  {Dullemond}}]{visser}
{Visser}, R., {van Dishoeck}, E.~F., {Doty}, S.~D., \& {Dullemond}, C.~P. 2009,
  \aap, 495, 881

\bibitem[{{Walsh} {et~al.}(2010){Walsh}, {Millar}, \& {Nomura}}]{walsh}
{Walsh}, C., {Millar}, T.~J., \& {Nomura}, H. 2010, \apj, 722, 1607

\bibitem[{{Willacy} \& {Woods}(2009)}]{will}
{Willacy}, K. \& {Woods}, P.~M. 2009, \apj, 703, 479

\bibitem[{{Woitke} {et~al.}(2009){Woitke}, {Kamp}, \& {Thi}}]{wkt}
{Woitke}, P., {Kamp}, I., \& {Thi}, W. 2009, \aap, 501, 383

\bibitem[{{Woodall} {et~al.}(2007){Woodall}, {Ag{\'u}ndez}, {Markwick-Kemper},
  \& {Millar}}]{umist}
{Woodall}, J., {Ag{\'u}ndez}, M., {Markwick-Kemper}, A.~J., \& {Millar}, T.~J.
  2007, \aap, 466, 1197

\end{thebibliography}

\end{document}